\documentclass[12pt,preprint,a4paper]{aastex}

\usepackage{graphics,epsf}
\usepackage{amsmath}                % American Mathematical Society package
\usepackage{amsfonts}               % American Mathematical Society fonts
\usepackage{amssymb}                % American Mathematical Society symbol
\usepackage{epsfig}                 % EPS figures
\def \cm{~\rm{cm}}
\def \s{~\rm{s}}
\def \km{~\rm{km}}

\def \K{~\rm{K}}
\def \g{~\rm{g}}

\def \erg{~\rm{erg}}

\def \yr{~\rm{yr}}

\def \pc{~\rm{pc}}
\def \kpc{~\rm{kpc}}

\def \keV{~\rm{keV}}

\begin{document}

\title{Transient outburst events from tidally disrupted asteroids near white dwarfs}

\author{Ealeal Bear$^{*1}$ and Noam Soker$^1$}

\affil{1. Department of Physics, Technion -- Israel Institute of
Technology, Haifa 32000, Israel;
$^{*}$ealeal@physics.technion.ac.il; soker@physics.technion.ac.il}

\begin{abstract}
We discuss the possibility of observing the transient formation event of an accretion disk from the tidal destruction process
of an asteroid near a white dwarf (WD). This scenario is commonly proposed as the explanation for dusty disks around WDs.
We find that the initial formation phase lasts for about a month and material that ends in a close orbit near the WD forms
a gaseous disk rather than a dusty disk. The mass and size of this gaseous accretion disk is very similar to that of Dwarf
Novae (DNe) in quiescence. The bolometric luminosity of the event at maximum is estimated to be $\sim 0.001-0.1L_\odot$.
Based on the similarity with DNe we expect that transient outburst events such as discussed here will be observed at
wavelengths ranging from visible to the X-ray, and be detected by present and future surveys.
\end{abstract}

Keywords: white dwarfs, planetary systems

% ==========================================================
\section{Introduction}
\label{sec:intro}
% ==========================================================

IR excess from white dwarfs (WDs, e.g. Becklin et al. 2005; Kilic et al. 2005, 2006; Jura et al. 2007a,b, 2009a;
Mullally et al. 2007; Kilic \& Redfield 2007; Jura 2008; Farihi et al. 2009, 2010a,b, 2011a, b;
Girven et al. 2011; Brinkworth et al. 2012) is commonly attributed to dusty disks formed from collisions of
small bodies with the WDs
(Zuckerman \& Becklin 1987; Graham et al. 1990; Brinkworth et al. 2009; Farihi et al. 2010a,b, 2011a;
 Debes et al. 2012).
Twenty five years after the first discovered dusty WD G$29-38$
(Zuckerman \& Becklin 1987), we know of the existence of more than
20 circumstellar dusty disks around WDs (e.g., von Hippel et al.
2007; Brinkworth et al. 2009; Farihi et al. 2010a, b ,2012; Dufour
et al. 2010; Melis et al. 2010, 2011; Girven et al. 2012; Kilic et
al. 2012; Xu \& Jura 2012). WDs that are heavily polluted with
metals might have accretion discs as well, even if no IR is
detected (Farihi et al. 2008, 2009, 2010a; Rafikov 2011; Debes et
al. 2012 and references therein). Farihi et al. (2009; see also
Kilic \& Redfield 2007) estimate that the fraction of DAZ WDs with
debris disks is $\sim {\rm few}~10\%$ for young WDs, with a
fraction of only few per cents at cooling age of $\sim 1~$Gyr.

Metals have short gravitational sedimentation time in WD photospheres (e.g., Koester 2009).
The ratio of the settling time to the cooling age of G149-28 and NLTT 43806, for example, is $\sim 10^{-7}$ (Zuckerman et al. 2011).
This implies that metal-polluted WDs, such as G238-44 (Zuckerman et al. 2003) and G$29-38$
(Koester et al. 1997; Zuckerman et al. 2003; Reach et al. 2009), must
be continuously supplied by fresh metals.

The composition of these WDs, as deduced from spectra, is
consistent with accretion from asteroids or comets, but not with
accretion from the ISM (e.g., Sion et al. 1990; Zuckerman et al.
2003; Jura 2003, 2006; Kilic \& Redfield 2007;
G{\"a}nsicke et al. 2007, 2008; Farihi et al. 2010a).
Furthermore, specific white dwarfs such as G$29-38$ were observed
with a spectral energy distribution that can be explained by
silicate compounds such as olivine (Reach et al. 2005). Since
comets cannot supply the amount of metals observed in some heavily
polluted WDs and are also dynamically less favored due to their
large distances from the star (Brinkworth et al. 2009), we will
discuss here only asteroid destruction.

Several studies calculate the mass residing around some WDs with IR excess, e.g., G$29-38$, to be $M_d\sim 10^{18}\g$,
which is the equivalent of a $\sim 5\km$ radius asteroid (Reach et al. 2005).
Kilic et al. (2012), however, argue that the masses of the disks in both G$29-38$
and PG 1541+651 are not well defined and might contain masses in the range of $10^{19}-10^{24}\g$ (Jura 2003;
Jura et al. 2007a, b, 2009b).
Interestingly, this mass range is similar to the disk mass of a typical dwarf nova (DN), an equality we will use later on.

The general consensus is that the loss of the asymptotic giant
branch (AGB) star envelope affects the stability of the planetary
system (Debes \& Sigurdsson 2002; Nordhaus et al. 2010; Veras et al.
2011; Bonsor et al. 2011), e.g., by increasing the orbital
separations of surviving planets. The change in orbits might lead to
instabilities that can perturbed the orbits of low mass bodies, such
as those in the asteroid belt, Kuiper belt, and the Oort cloud. Some
of these bodies collide later on with the WD remnant of the AGB
star. Tidal forces on the minor body destruct it and lead to the
formation of the dusty disk (e.g., Jura 2003; G{\"a}nsicke et al. 2006; Su et al. 2007; Metzger et al. 2012).
This is an optically thick disk located at $10R_{\rm WD}$ (e.g., Debes et al. 2011), where $R_{\rm WD}$ is the WD radius.

The concept of a collision of an asteroid with a compact remnant is not a new one
(e.g. Harwit $\&$ Salpeter 1973; Newman $\&$ Cox 1980a,b; Howard et al. 1981), and some studies examined
the initial flare, or transient event, that occurs during the process of asteroid destruction
(Soker et al. 2010; Campana et al. 2011; Zubovas et al. 2012 for destruction of an asteroid by Sgr A$^\ast$).
We suggest here that disks around WDs begin with a transient outburst event, and study the properties of such events.
We find that the increase in luminosity is not only important relative to
the luminosity of the WD, but also occurs at different wavelengths.
We are also motivated by the usage of tidal disruption of stars to detect inactive black holes
(e.g., Kasen \& Ramirez-Ruiz 2010) and the explanation of Campana et al. (2011) to the GRB 101225A
event as asteroid destructed by a neutron star (NS).

In Section 2 we discuss the mechanism that might induce the formation of such a transient outburst event and its luminosity.
In Section 3 we discuss some similarities of the proposed transient outburst event with properties of DNe.
Our short summary is in section 4.

% ==========================================================
\section{Peak luminosity}
\label{sec:Luminosity}
% ==========================================================

Once a body enters the tidal destruction radius $R_t$ it will disintegrate.
Part of the disintegrate material stays bound to the WD and creates an accretion disk, while
the rest of the mass escapes the system (e.g., Phinney 1989; Lodato \& Rossi 2011).
Evaporation by the WD radiation of the dust near the WD, at separations of $r \la 0.1 R_\odot$,
leads to the formation of a gaseous disk. G{\"a}nsicke et al. (2006) reported a metal-rich gas disk
around the moderately hot and young white dwarf ($T_{\rm eff} = 22020 \pm 200K$) SDSS1228$ + $1040.
The outer radius of the disk is $1.2 R_\odot$. Based on that, we expect that cooler WDs,
as typical for those with dusty disks, can evaporate dust to distances of $\sim {\rm few}\times 0.1 R_\odot$,
as appropriate for our proposed scenario.
The viscosity in the gaseous disk is large enough to allow rapid accretion of this material, leading to
the transient outburst event studied here.
Bound dust at a distance of $r \sim 1 R_\odot$ is accreted on a much longer timescale (Metzger et al. 2012),
and ensures that the WD photosphere stays polluted for a long time.

The tidal radius is given by (Davidsson 1999; Jura 2003)
\begin{equation}
R_t \simeq C_{\rm tide} R_{\rm WD} \left( \frac{\rho_{\rm WD}}{\rho_{\rm ast}} \right)^{\frac{1}{3}}
\simeq 1.3
\left(\frac{C_{\rm tide}}{2} \right)
\left(\frac{M_{\rm WD}}{0.6 M_\odot}\right)^{\frac{1}{3}} \left(\frac{\rho_{\rm ast}} {3 \g
\cm^{-3}}
\right)^{-\frac{1}{3}} R_\odot, % 1.169
\end{equation}
where $R_{\rm WD}$, $M_{\rm WD}$, and  $\rho_{\rm WD}$ are the radius, mass and density of the WD, respectively,
and $\rho_{\rm ast}$ is the density of the asteroid.
The value of the coefficient $C_{\rm tide}\simeq 1.3- 2.9$ dependents on orbital parameters of the asteroid, on its
rotation, and on its composition (Davidsson 1999; Jura 2003; Harris 1996).
We will used $C_{\rm tide}=2$, for a solid non-synchronized asteroid.

As the asteroid is tidally destructed, its material forms a very elongated structure along the orbit.
Bound material in the front falls back and collides with matter in the back of the `tail'.
A gaseous disk will be formed. There are two sources for the gas. First, the radiation from the WD will evaporate some of the
particles (dust) flowing close to it, $\sim {\rm few}\times 0.1 R_\odot$.
The second source is collisions of rubble-piles. The tidally destructed asteroid material stays in a very narrow plane.
Hence, there is a chance for some solid bodies, which are actually rubble-piles, to collide with each other. %%%%XXX
The exact probability will be worked out by numerical simulations, but it is not small because the trajectory of the different
asteroids fragments cross one another.

With a gaseous disk, we can workout the thermodynamics of the disk.
In a tidal destruction process about half of the mass is ejected and carries some extra energy.
This implies that the bound mass does not carry all its initial energy,
and its temperature will be somewhat less than the virial temperature
$T_{\rm vir} (r) \simeq 10^8 (r/0.1 R_\odot)^{-1} \K $ after it settles to a disk,
whether geometrically thin or thick.
The temperature of the disk can in principle be in the range $T \sim 0.001 T_{\rm vir} - T_{\rm vir} \simeq 10^5-10^8 \K$.
From the accretion time scale (eq. \ref{eq:tvisc1} below) and accretion rate (eq. \ref{Eq.dotMfb} below), we estimate the mass
in the disk within radius $\sim 0.1 R_\odot$ to be $\sim 10^{20} \g$. With an opacity of $\kappa > 1$ (as the mater contains mainly metals),
the disk is optically thick. The same conclusion can be reached by comparison with dwarf novae (see section \ref{sec:DN}).
The disk cannot be a geometrically thick disk.
This is because such a disk implies high temperature, which in turn implies a very high emission rate (because the disk is optically thick).
In addition, the high molecular weight of the metals also favors a thin disk.
If, for example, we take the disk scale hight to be 10 times the asteroid size, $H \sim 1000 \km$,
then at $r \sim 0.1 R_\odot$ we find $H/R \sim 0.01$.
Overall, at this preliminary study we can safely take $H/R \sim 0.01-0.1$. We scale quantities with $H/R =0.1$, but one should keep in mind
that a lower value is possible. We don't expect a thicker disk.

The gaseous disk will be formed very close to the WD, where we expect the dust particles to be evaporated.
Like Campana et al. (2011) we take the outer radius of the disk to be $\sim 2r_p$ (where $r_p$ is the periastron distance).
The scaled radius will be $R_d=r_p$.
The viscous timescale at radius $R_d$ in an accretion disk around a WD is (e.g., Dubus et al. 2001).
\begin{equation}
\begin{split}
t_{\rm{visc}} =\frac {R^2}{\nu}
\simeq 0.75
\left(\frac{\alpha}{0.1} \right)^{-1}
\left(\frac{R_d}{10 H }\right)
\left(\frac{v_\phi}{10 C_s}\right)
\left(\frac{M_{\rm WD}}{0.6M_\odot}\right)^{-\frac{1}{2}}
\left(\frac{R_d}{0.1R_\odot}\right)^{\frac{3}{2}} {\rm day},
\end{split}
\label{eq:tvisc1}
\end{equation}
where $\nu=\alpha C_s H$ is the viscosity of the disk parameterized with the disk $\alpha$ parameter, $H$ is the
vertical thickness of the disk (for a geometrically thin disk, $\frac{H}{R_d}<<1$),
$C_s$ is the sound speed and $v_\phi$ is the
Keplerian velocity. For our purpose the relevant radius $R_d$ is where the early fallback material concentrate which we take to be
the periastron distance of the destructed asteroid (Rosswog et al. 2004).

We note that Metzger et al. (2012) give a typical timescale of $\sim 2 \times 10^3 \yr$, which is
$\sim 10^6$ as long as the time we give here.
The reason is that they consider the formation of the gaseous disk from evaporation of dust at much later times.
They take the sound speed in the disk to be $\sim 2 \km \s^{-1}$, while we consider the violent formation of the
disk very close to the WD during the destruction process.
Their temperature comes from the typical evaporation temperature of dust (thousands of degrees Kelvin),
while our temperature comes from the velocity of the destructed asteroid $\sim 3000 \km \s^{-1}$,
which is equivalent to a temperature of $\sim 10^8 \K$.
We take $H\sim 0.1R_d$ and a sound speed of $\sim 100 \km \s^{-1}$.
Metzger et al. (2012) take $\alpha=0.001$ while we take $\alpha=0.1$.
These differences lead to the six order of magnitude difference in the estimated time scale.
The typical parameters we use here were used before.
Campana et al. (2011) use a very short viscous time scale for the accretion disk formed by a
destructed asteroid around a NS.
Soker et al. (2010) suggest that an inflated envelope of radius $\sim 0.5 R_\odot$ is formed from
the destruction process of the asteroid.
Our viscous time estimate and that of Metzger et al. (2012) are not in contradiction.
They apply to two different phases of the accretion process, at destruction and at a much later time, respectively.

The tidal disruption of stars that pass close to massive black holes has been studied in details
(e.g., Phinney 1989; Evans \& Kochanek 1989; Ulmer 1999; Ayal et al. 2000;
Gomboc \& {\v C}ade{\v z}; 2005Kosti{\'c} et al. 2009;
Lodato et al. 2009; Lodato \& Rossi 2011 and references therein).
Once a star is tidally disrupted half the mass becomes unbound and half the mass is bound.
The bound mass trajectory is highly eccentric and the most bound mass (has the lowest energy) returns within a
minimal time called the minimal fallback time and is given by (e.g., Ulmer 1999, Lodato \& Rossi 2011 and Giannios \& Metzger 2011 for recent papers on the subject)
\begin{equation}
t_{\rm min}=\frac{\pi}{2^{\frac{1}{2}}} \left( \frac{r_{p}}{R_{\rm ast}} \right)^{\frac{3}{2}}
\left( \frac{r_{p}^3}{G M_{\rm WD}} \right)^{\frac{1}{2}}
\simeq 31                                           % 30.718
\left(\frac{r_{p}}{0.1R_\odot}\right)^{3}
\left(\frac{R_{\rm ast}}{100 \km}\right)^{-\frac{3}{2}}
\left(\frac{M_{\rm WD}}{0.6 M_\odot} \right)^{-\frac{1}{2}} {\rm day}
\label{Eq.tmin}
\end{equation}

For the mass accretion rate we follow Campana et al. (2011) who applied it to an asteroid destructed around a NS,
and take
\begin{equation}
\dot M_{\rm FB}=\frac{M_{\rm ast}}{3 t_{\rm min}}
\left(\frac{t}{t_{\rm min}}\right)^{-\frac{5}{3}} \qquad {\rm for} \qquad t \ge t_{\rm min}.
\label{Eq.dotMfb}
\end{equation}
As the viscosity time is much shorter than the fall back time scale, shortly after the material fall back it is accreted.
The typical accretion rate at maximum is $\dot M_{\rm FB-max} \simeq 10^{15} (M_{\rm ast}/10^{22} \g) \g \s^{-1}$.

~From equations (\ref{Eq.tmin}) and (\ref{Eq.dotMfb}), and the expression for the potential on the surface of the WD
(assuming it is a slow rotator), we find the accretion luminosity after fall back has started ($t \ga t_{\rm min}$)
\begin{equation}
L_{\rm FB} \simeq 0.02 \left(\frac{\rho_{\rm ast}}{3 \g\cm^{-3}}
\right) \left(\frac{r_{p}}{0.1R_\odot}\right)^{-3}
\left(\frac{R_{\rm ast}}{100 \km}\right)^{\frac{9}{2}}
\left(\frac{M_{\rm WD}}{0.6 M_\odot} \right)^{\frac{3}{2}}
\left(\frac{R_{\rm WD}}{0.01 R_\odot} \right)^{-1}
\left(\frac{t}{t_{\rm min}}\right)^{-\frac{5}{3}}
L_\odot.
\label{Eq.Lfb}
\end{equation}
The mass of the asteroid with the scaling above is $\sim 10^{22} \g$.
The mass can be larger, as the mass of the asteroid that is assumed to have collided with the most polluted WD
is assumed to be $\sim 10^{24}\g$ (Jura 2006), similar to the mass of Ceres.
We also note that a viscosity time scale 100 times as large as assumed here in equation (\ref{eq:tvisc1}) would
increase the accretion scale from one to about two months, and will reduce the peak luminosity by factor of few only.

The sharp increase in luminosity and the rapid decline forms a transient outburst event.
Soker et al. (2010) suggest that an inflated envelope of radius $R_e \sim 0.5 R_\odot$ is formed.
We take the envelope mass to be $\eta M_{\rm ast}$, and the effective radius for the calculation of its gravitational
energy to be $R_e \simeq 0.1 R_\odot \simeq 10 R_{\rm WD}$.
The thermal (Kelvin-Helmholtz) timescale of this envelope is of the order of
\begin{equation}
t_{\rm KH} \simeq \frac{G M_{\rm WD} \eta M_{\rm ast}}{L_{\rm FB-max} R_e} \simeq
0.3 \eta t_{\rm min},
\label{Eq.tkh}
\end{equation}
where in the second equality we used equation (\ref{Eq.tmin}).
The thermal time scale is much shorter than the accretion timescale. The violent tidal destruction process might lead to
a temporary inflated envelope around the WD.
To check whether such a temporary envelope is formed requires 3D numerical simulations with accurate radiative transfer calculation.

The luminosity is very sensitive to the asteroid radius $R_{\rm ast}$ and the periastron distance $r_{p}$.
Long-lived IR disks observed around WD usually have an outer disk radii of $\sim R_\odot$ (Jura 2003).
We, however, are interested in the gaseous disk that is formed close to the WD immediately after the destruction of the steroid,
a time scale of $\la {\rm few} \times t_{\rm min}$.
For that we take the disk radius to be $R_d \sim 0.1R_\odot$.
The tidal destruction radius limits the periastron distance $r_p \la 1R_\odot$.
Considering this limit and the mass that is required to pollute WDs and form the IR excess, flares with
peak luminosities of $L \ga 0.02 L_\odot \simeq 10^{32} \erg \s^{-1}$ are expected to occur
at the formation of the dusty disks around WDs. A large part of the bound material will stay in the form of dust particles.
We expect therefore that the decline in luminosity will become more rapid with time that $t^{-5/3}$,
and the burst will practically last for a time of $\sim t_{\rm min}$.
Less than half the bound mass will be accreted during this phase.

% ==========================================================
\section{Some similarities with dwarf novae}
\label{sec:DN}
% ==========================================================
Dwarf novae (DNe) involve a primary star which is a WD that accretes mass from an accretion disk.
Occasionally the WD experiences outbursts due to enhanced mass accretion rate.
The secondary that supplies the mass is usually a low mass main sequence star.
These outbursts are driven by an instability of the disk that enhances mass accretion rate (e.g., Frank et al. 1992).
In quiescence DNe emit hard X-ray from the optically thin boundary layer (BL) formed between the optically thick disk and the WD
surface (e.g., Szkody et al. 2002; Ishida 2010).

Pandel et al. (2005), for example, study eight DNe that were observed by the XMM Newton in
quiescence with typical X-ray luminosity of $10^{31} \erg\s^{-1} - 6.6\times 10^{32} \erg\s^{-1}$
and deduced accretion rates of $6 \times10^{13} \g \s^{-1}- 6 \times 10^{15} \g \s^{-1}$ and a typical BL
temperatures of $T_{\rm BL} \simeq 8-55 \keV$.
The accretion disk luminosity might be mainly in the UV, and its luminosity is expected to
be similar to that of the BL  (e.g., Pringle \& Savonije 1979; Balman et al. 2011).
Godon \& Sion (2005) argue that in some cases $\sim 20\%$ of the BL luminosity is in the X-ray, and the rest is in the UV.
The disk behavior in DNe might be much more complicated than what we consider here, e.g.,
due to instabilities and WD-magnetic fields truncation of the disk (e.g., Lasota 2001).
However, we compare the asteroid destruction flares with DNe behavior to warn against confusion in observing these events.

DNe within $0.5\kpc$ and can be well observed in X-ray (Ulman \& Sion 2006).
This further motivates us to consider the similarities between DNe and the proposed transient outburst events, as the later
are well within detection capabilities.
The DN VW Hyi, for example, has a typical accretion rate onto the WD of
$5\times 10^{-12}M_\odot \yr^{-1} \sim 3\times10^{14}\g\s^{-1}$ (Pandel et al. 2003), and its
X-ray luminosity is $\sim 8\times 10^{30}\erg\s^{-1}$.
The accretion rate from many tidally disrupted asteroid is expected to be $\ga 10^{14}\g$, resulting in
detectable X-ray, UV, and visible bands emission.

We note that the comprehensive thermal-viscous disc instability model (DIM, Lasota 2001)
suggests very cold disks ($ \sim 5800K$) which are neutral and therefore no X - ray emission is expected.
This does not agree with observations (Lasota 2001 and references therein).
The size of the disks in DNe is similar to the size of the disk when an asteroid is tidally disrupted near a WD.
The inner disk radius might not extend up to the WD surface as suggested by Lasota (2001) due to truncation caused by
the magnetic field.
The X-ray luminosity for the optically thin BL case was calculated by
Patterson \& Raymond (1985) for the spectral range of 0.2-4\keV.
As the accretion rate expected in the proposed transient outburst events is in most cases $<10^{16} \g \s^{-1}$,
we can take the BL to be optically thin, and use results for DNe in quiescence.
In table 1 we compare some properties of DNe with those of the proposed transient outburst events.
\begin{table}
Table 1: Comparing dwarf novae with asteroid transient outburst events

\bigskip
\begin{tabular}{|l|c|c|}
\hline
Property & Dwarf Novae in quiescence & Tidal disruption of asteroid \\
\hline
$L_{\rm visible}$ ($L_\odot$) & $\sim 0.1$$^{(l)}$  & $\sim 0.01^{(m)}$ \\

\hline $L_{\rm X}$ $({\rm erg}\s^{-1})$ &  $10^{30} {\rm erg}\s^{-1} -
10^{33} {\rm erg}\s^{-1}$ $^{(a,d)}$ &${\rm few} \times 10^{31} {\rm erg}\s^{-1}$ \\
\hline
Composition & Hydrogen, Helium  and metals$^{(e)}$ & Mainly metals$^{(e)}$ \\
\hline
$R_{\rm disk}[{\rm Outer~Radius}]$ & $\sim 0.2R_\odot$$^{(f,g,j)}$  & $\sim 0.1R_\odot$ $^{(b,h)}$ \\
\hline
$M_{\rm disk}$ & $10^{20}- 10^{21}\g$ $^{(i, k)}$ & $\frac{1}{2}\times M_{\rm ast}$$^{(b)}$ \\
\hline Mass accretion rate $(\g\s^{-1})$ &$6
\times10^{13} \g \s^{-1}- 6 \times 10^{15} \g \s^{-1}$$^{(a,c)}$& $10^{13}-10^{16}\g \s^{-1}$$^{(b)}$ \\
\hline
\end{tabular}

\footnotesize
\bigskip
Notes:

        (a) Balman et al. (2011).\normalsize

        (b) $M_{\rm ast}\sim 10^{20}-10^{24}\g$, where we used $M_{\rm ast}\sim 10^{22}\g$ (for more details see section \ref{sec:Luminosity}).\normalsize

        (c) Long et al. (2005).\normalsize

        (d) Medvedev \& Menou (2002).\normalsize

        (e) See section \ref{sec:DN}.\normalsize

        (f) Smak (2002).\normalsize

        (g) Rafikov (2011).\normalsize

        (h) G{\"a}nsicke et al. (2006); Brinkworth et al. (2009).\normalsize

        (i) Godon \& Sion (2005) and references therein.\normalsize

        (j) Truss et al. (2004) and references therein.\normalsize

        (k) Frank et al. (1992).\normalsize

        (l) We note that a significant contribution might come from the WD (Patterson 2011;
            Godon \& Sion 2005) .\normalsize

        (m) In DNe the WD contributes significantly to the luminosity, while in many tidal destructed asteroid
            events the WD is cooler and does not contribute much to the luminosity (Koester et al. 2011).
            In cases where the contribution of the WD is important
            (e.g., G{\"a}nsicke et al. 2006, {{{ 2007, 2008 }}}; Townsley
            \& G{\"a}nsicke 2009) it is constant and does not
            change over a timescale of the transient event formation.
            (few weeks, Koester \& Wilken 2006; Girven et al. 2012).

       \end{table}

We compare the transient outburst event with DNe in quiescence as the BL is expected to be optically thin. In
the transient outburst event the luminosity will decrease and fade within several weeks.
The time scale is dictated by the fall back time.
In DNe the BL is optically thick during outburst, and hence we did not compare the transient outburst event to
DN outbursts.
The DN outburst is dictated by viscosity in the disk, and usually lasts for several days.
Superoutbursts in DNe can last for few weeks.
In case of a massive asteroid with a small periastron distance $r_p$, accretion rate can be very
high in the first several days. In that case the transient formation event will be more similar to DN outbursts.

As with DNe, non-negligible fraction of the emission will be in the optical band, and optical surveys
from the ground can detect these transient outburst events.

There are some differences between DNe and the WD-asteroid transient outburst events.
\newline
(1) In DNe most FUV radiation comes from the hot WD (e.g., Godon at al. 2008).
In the transient outburst events in most (but not all) cases the WD is somewhat cooler, and the FUV will come from the inner part of the disk.
{{{\newline}}}
(2) In our scenario the material is rich in metals, that at the relevant temperatures of ${\rm few} \times 10^4 \K$
are ionized mostly twice. The molecular weight is $10-30$ times as high as that of ionized solar composition gas,
and the sound speed will be $3-5$ times lower. The evaporation rate is proportional to the
sound speed (e.g., Meyer \& Meyer-Hofmeister 1994).
As no complete theory of evaporation exist (Lasota 2001) we cannot say more beside expecting
lower evaporation rate than in DNe.
\newline
{{{ (3) }}} No He lines are expected in the transient outburst event. Hydrogen will
show weaker lines. Some metals with relatively strong spectral
lines in DNe, e.g., nitrogen, will be absent, or have weak lines,
in transient outburst events. Other metals will show stronger line in the
transient outburst events, such as carbon, oxygen, and iron depending on the asteroid composition {{{ {(Jura et al. 2012).} }}}
Classification of asteroids is complicated and is strongly
connected to their composition. For our purpose we will focus on
the three main types of asteroids (Lupishko \& Belskaya 1989;
Wiegert et al. 2007; Mainzer et al. 2011): \begin{itemize}
\item C-type asteroids (carbonaceous) are considered to have a composition similar to the Sun and are depleted in hydrogen,
helium and other volatiles. They host $75\%$ of our asteroid belt. We would like to emphasis that although these asteroids
contain relative large amounts of carbon compared to solar system rocks, they are still depleted in carbon with
respect to solar composition (Lee et al. 2010).
\item S-type (silicaceous) are composed of metallic iron mixed
with iron and magnesium-silicates. They host the inner part of the asteroid belt and account for $17\%$ of our
asteroid belt.
\item
M-type (metallic) are mainly composed of metallic iron. However, from the surface composition of M-type
 asteroids it is evident that they are not composed solely from metallic iron, but rather contain
 considerable amounts of silicate-component (like stony-iron and enstatite chondrite meteorites).
\end{itemize}

We note that the typical asteroid mass inferred from the convection zones of many polluted
He WDs is larger than the typical mass of the solar system asteroids.
Hence, the estimated mass and composition of asteroids in our solar system might not be representative,
and should only serve as a rough estimate
(Dufour et al. 2012; Giammichele et al. 2012).
Since the WD is typically cooler and the source of mass is an asteroid, molecules are expected to reside in the
vicinity of the WD in these transient outburst events.
This can lead to the presence of the C$_2$ swan bands in transient formation events (Soker et al. 2010).
The swan bands are observed in DQ WDs.
DQ WDs are cold ($T_{\rm eff} < 13000K$) WDs with a helium-dominated atmosphere that are
enriched in carbon that was dredged-up from the core (by the deep helium convection zone;
Vornanen et al. 2010; Kowalski 2010 and references therein).
 Although swan band are usually observed at temperatures of $5000 - 6000\K$ and generally weaken
with increasing temperature, in a carbon-dominated atmosphere they would be extremely strong even at $10,000 \K$
(Dufour et al. 2005; Koester \& Knist 2006; G{\"a}nsicke et al. 2009; Soker et al. 2010).
The molecule will not be formed near the boundary layer, but rather at the surface of the disk, where the
temperature allows the formation of the molecules.
As C-type asteroid, that are rich in carbon, are thought to be the most common type in the
Main Belt (Mainzer et al. 2012), it is possible that C$_2$ swan bands be detected in some transient outburst events.

% ==========================================================
\section{Summary}
\label{sec:summary}
% ==========================================================

We examined the possibility that the tidal destruction process of an asteroid near a WD will form a transient outburst
event that can be detected. We considered an asteroid with a periastron distance larger than the
radius of the WD but smaller than the tidal destruction radius.
Namely, we did not consider direct hit of the WD. For such a case the asteroid is tidally
destructed in a way that about a half of the mass escapes while the rest stays bound. We considered the bound material with
the lower most energy. We expect that $\sim 10\%-20\%$ of the asteroid mass that is most bound (have lowest energy) will suffer
strong dissipation near the WD and will form a gaseous disk on the fallback timescale $t_{\rm min}$ (eq. \ref{Eq.tmin}).
The viscous timescale of this material is of the order of the fallback time or shorter (eq. \ref{eq:tvisc1}).

This most bound material is accreted at a peak rate of
$\dot M_{\rm FB-max} \simeq 10^{15} (M_{\rm ast}/10^{22} \g) \g \s^{-1}$
(eq. \ref{Eq.dotMfb}), where $M_{\rm ast}$ is the asteroid mass.
Accretion starts on a time scale $t_{\rm min}$ of days to months, strongly depending on the periastron
distance and the asteroid radius (eq. \ref{Eq.tmin}).
The accretion rate over that period can be $\sim 10^{13}-10^{16}\g\s^{-1}$ (eq. \ref{Eq.dotMfb}).
The total luminosity can be $\sim 0.02L_\odot$ from visible to X-ray.

This accretion rate and luminosity are similar to those of dwarf novae (DNe) in quiescence, and we compare their properties
in Table 1. However, differences exist as we elaborate in Sec \ref{sec:DN}. The two main differences are:

(1) In DNe (but not in all transient outburst events) the WD may be hot and luminous.

(2) No He lines are expected in transient outburst events. Hydrogen will
show weaker lines. Because of the differences in the composition of asteroids and the secondaries in DNe,
there will be differences in metal lines relative intensities as well.

We note that the duration of the event, about a months, is of the order of the duration of
DN cycles (Lasota 2001 and references therein). Taking at face values, this implies that disk instabilities
might cause large variations in the luminosity even during the several weeks long transient event.
The exact mechanism in our case is different though, because the ionization potential of some metals is lower
than that of the hydrogen, and instabilities might occur at different conditions.

With more and more large surveys that are being conducted, such as the Large Synoptic Survey Telescope (LSST),
we expect that such events will be reported in the near future.
The solar-vicinity density of WDs is estimated as $\sim 5\times 10^{-3} M_\odot \pc^{-3}$ (Mendez \& Minniti 2000; Holberg et al. 2002),
 implying that
there are ${\rm {few}} \times 10^6 $ WDs within $1\kpc$ from the Sun.
The number of WDs with derbies disks depends on the WD cooling age (Farihi et al. 2009),
therefore we take $1\%$ as an average estimation.
Assuming that $\sim 1\%$ of all WDs have disks, we find that the number of WDs with disks within
$1 \kpc$ is $N_{D1} \sim 10^4$. The disk lifetime is
estimated as $\tau_D \simeq 3\times 10^4 - 5 \times 10^6 \yr$ (Girven et al. 2012).
From these values we crudely estimate the event rate that is bright enough to be
detected in surveys to be  $N_{\rm event} \sim \frac{ N_{D1}}{\tau_D } \sim 0.002-0.3 \yr^{-1}$.
Large asteroids lead to much brighter events that can be detected beyond $1 \kpc$.
The rate can be increased is we consider cases where no dusty disks are formed.
A comet is tidally destructed, but it forms a gaseous disk such that no IR excess is observed.
However, an initial flare of the type studied here might be detected.
Overall, the total rate of such detected events might be $\sim 0.01-1 \yr^{-1}$.

One type of places where such transient outburst events are likely to be detected is open clusters.
We suggest that the transient outburst events studied here might occur in open clusters,
where neighboring stars can perturbed the planetary system of newly formed WDs.
In open clusters metallicity is high and we expect a high frequency of planetary systems.
The Orion nebula, for example, is at a distance of only $412 \pc$;
it is very likely that if we can observe hard X ray from DNe (see section \ref{sec:DN}) at quiescence at this distance,
we will be able to observe these transient outburst events near WDs in Orion.

It should be noted that Kilic et al. (2012) reported the observation of the disk around
PG1541, which is only $55\pc$ away from the Sun.
This observation demonstrates that although extensive surveys have been conducted,
such as WISE InfraRed Excesses around
Degenerates (WIRED) and Spitzer (Debes et al. 2011; Farihi et al. 2012) to distance of
$100\pc$, and $1000\pc$,  respectively, our knowledge of the nearby dusty
WDs population is far from complete.
We encourage analysis of large surveys to be alert to such events.

We thank Jay Farihi, Patrick Godon, Boris G{\"a}nsicke, Michael Jura, Sergio Campana, Eran Ofek, Jean-Pierre Lasota,
Roman Rafikov, Mukremin Kilic, and the two referees, for helpful tips and comments.
This research was supported by the Asher Fund for
Space Research at the Technion, The US - Israel Binational Science Foundation,
and the Israel Science Foundation.

% %%%%%%%%%%%%%%%%%%%%%%%%%%%%%%%%%%%%%%%%%%%%%%%%%%%%%%%%%%%%%%%%%%%%%%%%%%%%%%%%%%%%%
% %%%%%%%%%%%%Refrences
% %%%%%%%%%%%%%%%%%%%%%%%%%%%%%%%%%%%%%%%%%%%%%%%%%%%%%%%%%%%%%%%%%%%%%%%%%%%%%%%%%%%%%

\end{document}